\documentclass[12pt]{article}
\usepackage{graphicx}
\usepackage{epsfig}
\usepackage{latexsym}
\usepackage{latexsym}
\usepackage{amssymb}
\usepackage{amsmath}

\begin{document}
\author{Khireddine Nouicer\footnote%
{Permanent address: Laboratory of Theoretical Physics, Faculty of
Sciences, University of Jijel, BP 98 Ouled Aissa, 18000 Jijel,
Algeria.}\\
Frankfurt Institute of Advanced Studies,\\
John Wolfgang Goethe Universit\"at,\\
Ruth-Moufang-Str. 1, 60438 Frankfurt am Main, Germany\\
{Email: nouicer@fias.uni-frankfurt.de}}
\title{Entropy of the
Randall-Sundrum black brane world to all orders in the Planck length }
\date{}
\maketitle
\begin{abstract}
 We study the effects, to all orders
in the Planck length from a  generalized uncertainty principle
(GUP), on the statistical entropy of massive scalar bulk fields
in the Randall-Sundrum black brane world. We show that the Bekenstein-Hawking area law is not preserved,  and contains small corrections terms proportional to the black hole inverse area.
\end{abstract}

\section{Introduction}

The possibility of existence of extra dimensions has opened exciting and
promising ways to investigate phenomenological and cosmological aspects of
quantum gravity. Models with extra dimensions and an effective fundamental
scale of the order of the TeV have been proposed as possible solution to the
gauge hierarchy problem \cite{arkani1}-\cite{sundrum2}. Particularly, the
Randall-Sundrum models \cite{sundrum1,sundrum2} have attracted a great
attention and their cosmological implications intensively studied \cite%
{rizzo}-\cite{20}. On the other hand, since the seminal works of Bekenstein \cite%
{beken} and Hawking \cite{hawking}, the computation of the entropy of a
black hole remains an active field of research. Various approaches and
methods have been employed. Among them, the brick-wall method \cite{brick},
which is a semi-classical approach, has been applied to various BH
geometries \cite{medved} (and references therein). However, this approach
suffers from the implementation of unnatural arbitrary ultraviolet and
infrared cutoffs. Recently, with the advent of  generalized uncertainty
principles (GUPs), originating from several studies in string theory
approach to quantum gravity \cite{vene}-\cite{koni}, loop quantum gravity
\cite{Garay}, noncommutative space-time algebra \cite{mag01}-\cite{mangano}
and black holes gedanken experiments \cite{mag02,scar}, the
contribution to the entropy of quantum states with momentum above a given
scale has been suppressed and the UV divergence completely removed (see \cite%
{nouicer01} for an extensive list of references).

Recently, the calculation of the statistical entropy of thermal bulk
massive scalar fields on the Randall-Sundrum brane background has
been performed with a GUP to leading order in the Planck length
\cite{kim}, and the
effect of the GUP has been only considered on the 3-brane. On the hand a careful analysis of the entropy near the horizon to all orders in the Planck length has been performed for the (3+1)-dimensional Schwarzschild black hole \cite{park} and for the 2+1)-dimensional de Sitter black hole \cite{yong}.In this
paper, we extend this calculation to all orders in the Planck
length, and consider the regularizing effect of the GUP, first on the full volume of the space-time,
and then on the brane. In section 2, we introduce a version of the
GUP containing gravitational corrections to all orders in the Planck
length, and investigate some of its quantum implications. In section
3, we obtain a novel equation of states of density for the extra and radial modes. In section 4, using the near horizon geometry approximation and considering the effect of the GUP on the bulk states, we derive the free energy
of a massive bulk scalar field and by means of the first law of
thermodynamics we obtain the
GUP-corrected Bekentein-Hawking area law for the entropy . Then, in order to compare our results with obtained by the brick-wall method and with the
GUP to leading order in the Planck length, we ignore the effect of the GUP on the extra direction states density and
compute again the free energy and the entropy. The last section is
devoted to a summary and a discussion of the results obtained.

\section{Generalized uncertainty principle (GUP)}

One of the most interesting consequences of all promising quantum
gravity candidates is the existence of a minimal observable length
on the order of the Planck length.
The idea of a minimal length can be modeled in terms of a quantized
space-time and goes back to the early days of quantum field theory \cite%
{snyder} (see also $\cite{connes}-\cite{bondia}$ ). An alternative approach is to
consider deformations to the standard Heisenberg algebra \cite{kempf,mangano}%
, which lead to  generalized uncertainty principles. In this section we
follow the latter approach and exploit results recently obtained. Indeed, it
has been shown in the context of canonical noncommutative field theory in
the coherent states representation \cite{spallucci} and field theory on
non-anticommutative superspace \cite{Moffat,nouicer}, that the Feynman
propagator displays an exponential UV cut-off of the form $\exp \left( -\eta
p^{2}\right) $, where the parameter $\eta $ is related to the minimal
length. This framework has been further applied, in series of papers \cite%
{nicolini}, to the black hole evaporation process.

At the quantum mechanical level, the essence of the UV finiteness of the
Feynman propagator can be also captured by a non linear relation, $k=f(p)$,
between the momentum and the wave vector of the particle \cite{Sabine}. This
relation must be invertible and has to fulfil the following requirements:

\begin{enumerate}
\item For  energies much smaller than the cut-off the usual dispersion relation is
recovered.

\item The wave vector is bounded by the cut-off.
\end{enumerate}

In this picture, the usual commutator between the commuting position and momentum operators is generalized to
\begin{equation}
[X,P]=i\hbar\frac{\partial p}{\partial k}\Leftrightarrow \Delta X\Delta P\geq\frac{\hbar}{2}\left|\left\langle\frac{\partial p}{\partial k}\right\rangle\right|,
\end{equation}
and the momentum measure $d^{n}p$ is deformed as $%
d^{n}p\prod_{i}\frac{\partial k_{i}}{\partial p_{j}}$. In the
following, we will restrict ourselves to the isotropic case in one
space-like
dimension. Following \cite{spallucci,nouicer} and setting $\eta =\frac{%
\alpha L_{Pl}^{2}}{\hbar ^{2}}$ we have
\begin{equation}
\frac{\partial p}{\partial k}=\hbar  { \exp}\left( \frac{\alpha L_{Pl}^{2}%
}{\hbar ^{2}}p^{2}\right) ,  \label{measure}
\end{equation}%
where $\alpha $ is a dimensionless constant of order one.

From Eq.$\left( \ref{measure}\right) $ we obtain the dispersion relation
\begin{equation}
k\left( p\right) =\frac{\sqrt{\pi }}{2\sqrt{\alpha} L_{Pl}} {  erf}\left( \frac{%
\sqrt{\alpha }L_{Pl}}{\hbar }p\right) \label{mdr},
\end{equation}%
from which we have the following minimum Compton wavelength
\begin{equation}
\lambda _{0}=4\sqrt{\pi \alpha }L_{Pl}.  \label{compton}
\end{equation}
We note that a dispersion relation similar to the one given by
Eq.(3) has been used
 recently to investigate the effect of the
minimal length on the running gauge couplings \cite{sab}. In the
context of trans-Plankian physics, modified dispersion relations
have been also used to study the spectrum of the cosmological
fluctuations. A particular class of MDRs frequently used in the
literature \cite{slot, jora} is the well known Unruh dispersion
relations given by $k(p) = tanh^{1/\gamma}(p^{\gamma})$, with
$\gamma$ being some positive integer \cite{unruh}.

Let us show that the above results can be obtained from the
following momentum space representation of the position and momentum
operators \
\begin{equation}
X=i\hbar \exp \left( \frac{\alpha L_{Pl}^{2}}{\hbar ^{2}}P^{2}\right) {%
\partial _{p}}\qquad P=p.  \label{xp}
\end{equation}
The corrections to the standard Heisenberg algebra become effective in the
so-called quantum regime where the momentum and length scales are of the
order of the Planck mass $M_{Pl}$ and  the Planck length $L_{Pl}$
respectively.

The hermiticity condition of the position operator implies  modified completeness relation and modified scalar product given by
\begin{equation}
\int dpe^{-\frac{\alpha L_{Pl}^{2}}{\hbar ^{2}}p^{2}}|p\rangle \langle p|=1
\label{ferm}
\end{equation}%
\begin{equation}
\left\langle p\right\vert \left. p^{\prime }\right\rangle =e^{\frac{\alpha
L_{ {Pl}}^{2}}{\hbar ^{2}}p^{2}
}\delta \left( p-p^{\prime }\right) .
\end{equation}%
From Eq.$\left( \ref{ferm}\right) $, we observe that we have reproduced the
Gaussian damping factor in the Feynman propagator \cite{spallucci,nouicer}.

The algebra defined by Eq. $\left( \ref{xp}\right) $ leads to the following
generalized commutator and  generalized uncertainty principle (GUP)
\begin{equation}
\left[ X,P\right] =i\hbar \exp \left( \frac{\alpha L_{Pl}^{2}}{\hbar ^{2}}%
P^{2}\right) ,\quad \left( \delta X\right) \left( \delta P\right) \geq \frac{%
\hbar }{2}\left\langle \exp \left( \frac{\alpha L_{Pl}^{2}}{\hbar ^{2}}%
P^{2}\right) \right\rangle .  \label{GUP}
\end{equation}%

In order to investigate the quantum mechanical implications of this
deformed algebra, we solve the relation $\left(\ref{GUP}\right)$ for
$\left( \delta P\right) $ with the equality. Using the property
$\left\langle P^{2n}\right\rangle \geq \left\langle
P^{2}\right\rangle $ and $\left( \delta P\right) ^{2}=\left\langle
P^{2}\right\rangle -\left\langle P\right\rangle ^{2}$, the
generalized uncertainty relation is written as
\begin{equation}
\left( \delta X\right) \left( \delta P\right) =\frac{\hbar }{2}\exp \left(
\frac{\alpha L_{P {l}}^{2}}{\hbar ^{2}}\left( \left( \delta P\right)
^{2}+\left\langle  P\right\rangle ^{2}\right) \right) .
\end{equation}%
Taking the square of this expression we obtain
\begin{equation}
W\left( u\right) e^{W\left( u\right) }=u,  \label{lam},
\end{equation}%
where we have set $W(u)=-2\frac{\alpha L_{Pl}^{2}}{\hbar ^{2}}\left( \delta
P\right) ^{2}$ and $u=-\frac{\alpha L_{Pl}^{2}}{2\left( \delta X\right) ^{2}}%
e^{-2\frac{\alpha L_{P {l}}^{2}}{\hbar ^{2}}\left\langle P\right\rangle
^{2}}.$

The equation given by Eq.$\left(\ref{lam}\right)$ is exactly the definition
of the Lambert function \cite{Lambert}, which is a
multi-valued function. Its different branches, $W_k(u)$, are labeled by the integer $%
k=0,\pm 1,\pm 2,\cdots $. When $u$ is a real number Eq.$\left(\ref{lam}%
\right)$ have two real solutions for $0\geq u\geq -\frac{1}{e}$, denoted by $%
W_{0}(u)$ and $W_{-1}(u)$, or it can have only one real solution for $u\geq
0 $, namely $W_{0}(u)$ . For -$\infty <u<-\frac{1}{e}$, Eq.(\ref{lam}) have
no real solutions.

Finally, the momentum uncertainty is given by
\begin{equation}
\left( \delta P\right) =\frac{\hbar}{\sqrt{2\alpha}L_{Pl}}\left(-W\left(-\frac{\alpha L_{Pl}^{2}}{2\left( \delta X\right) ^{2}}%
e^{-2\frac{\alpha L_{P {l}}^{2}}{\hbar ^{2}}\left\langle P\right\rangle
^{2}}\right)\right)^{1/2}.  \label{argu}
\end{equation}%
From the argument of the Lambert function we have the following condition
\begin{equation}
\frac{\alpha L_{Pl}^{2}e^{\frac{2\alpha L_{Pl}^{2}}{\hbar ^{2}}%
\left\langle P\right\rangle ^{2}}}{2\left( \delta X\right) ^{2}}\leqslant
\frac{1}{e},
\end{equation}%
which leads to a minimal uncertainty in position given by
\begin{equation}
\left( \delta X\right) _{\min }=\sqrt{\frac{e\alpha}{2}} L_{Pl}e^{\frac{%
\alpha L_{Pl}^{2}}{\hbar ^{2}}\left\langle P\right\rangle ^{2}}.
\end{equation}%
The absolutely smallest uncertainty in position or minimal length \ is
obtained for physical states for which we have $\left\langle P\right\rangle
=0$ and $\left( \delta P\right) =\hbar /\left( \sqrt{2\alpha }L_{P {l}%
}\right) ,$ and is given by
\begin{equation}
\left( \delta X\right) _{0}=\sqrt{\frac{\alpha e}{2}}L_{Pl}.  \label{min}
\end{equation}%
In terms of the minimal length the momentum uncertainty becomes
\begin{equation}
\left( \delta P\right) =\frac{\hbar\sqrt{e}}{2(\delta X)_0}\left( -%
W\left( -\frac{1}{e}\left( \frac{(\delta X)_{0}}{(\delta X)}%
\right) ^{2}\right) \right)^{1/2} .  \label{argup}
\end{equation}%
This equation can be inverted to obtain the position uncertainty as
\begin{equation}
\left(\delta X\right)=\frac{\hbar}{2\left(\delta P\right)}\hbox{exp}\left(\frac{4\left(\delta X\right)_0^2}{\hbar^2 e}\left(\delta P\right)^2\right).
\end{equation}
In figure 1, we show the variation of the $\delta X$ with $\delta P$. The minimum corresponds to the location of the maximal localization states for which $\langle X\rangle=\xi$ and $\langle P\rangle=0$. We observe that for $\alpha$ large, corresponding to strong gravitational field, the uncertainty on the momentum operators becomes bounded, which is not the case in the standard situation with Heisenberg uncertainty principle (HUP) ($\alpha\longrightarrow 0$).
\begin{center}
\includegraphics[height=8cm,
width=10cm]{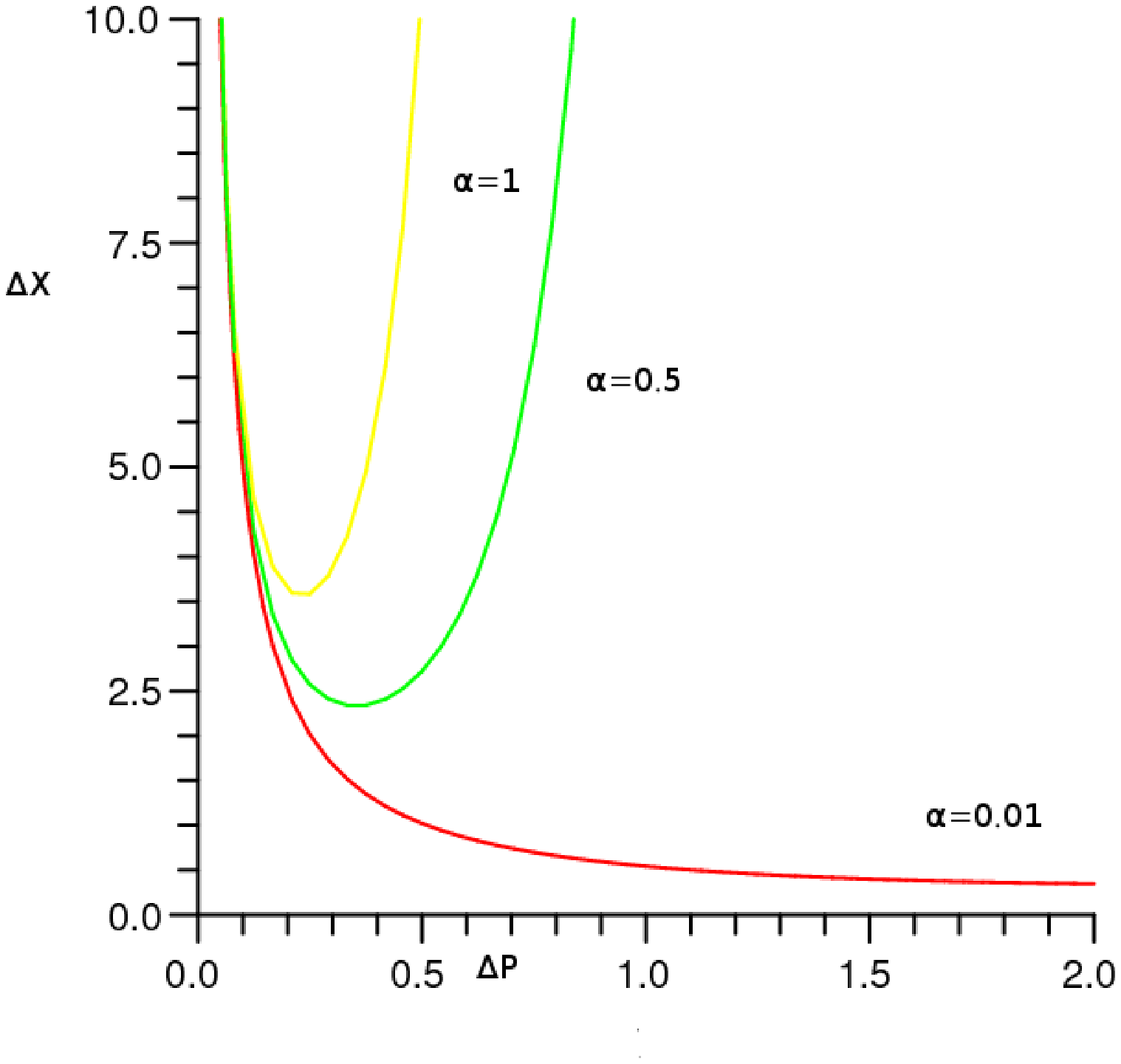}
\end{center}
\begin{center}
{Figure 1: Generalized uncertainty relation
.}
\end{center}
Let us observe that $\frac{1}{e}\frac{(\delta X)_{0}}{(\delta
X)}<1$ is a small parameter by virtue of the GUP, and then perturbative
expansions to all orders in the Planck length can be safely
performed.

Indeed, a series expansion of Eq.(\ref{argup}) gives the corrections to the
standard Heisenberg principle
\begin{equation}
\delta P\simeq \frac{\hbar }{2\left( \delta X\right) }\bigg( 1+\frac{1}{2e}%
\left( \frac{(\delta X)_{0}}{(\delta X)}\right) ^{2}+\frac{5}{8e^{2}}\left(
\frac{(\delta X)_{0}}{(\delta X)}\right) ^{4}
+\frac{49}{48e^{3}}\left( \frac{%
(\delta X)_{0}}{(\delta X)}\right) ^{6}+\ldots \bigg) .
\end{equation}%
This expression of $\left( \delta P\right) $ containing only odd powers of $%
\left( \delta X\right) $ is consistent with a recent analysis in which
string theory and loop quantum gravity, considered as the most serious
candidates for a theory of quantum gravity, put severe constraints on the
possible forms of GUPs and MDRs \cite{nozari}.

Before ending this section, we briefly recall the form of the GUP to leading
order in the Planck length, recently used by Kim et al. \cite{kim}. This GUP
is given by
\begin{equation}
\left( \delta X\right) \left( \delta P\right) \geq \frac{\hbar }{2}\left( 1+%
\frac{\alpha L_{Pl}^{2}}{\hbar ^{2}}\left( \delta P\right) ^{2}\right) .
\label{logup}
\end{equation}%
A simple calculation leads to the following minimal length
\begin{equation}
\left( \delta X\right) _{0}=\sqrt{\alpha }L_{Pl},
\end{equation}%
which is of order of the Planck length. However, as nicely noted in \cite%
{Sabine}, this form of the GUP do not fulfil the second requirement listed
above.

In the following sections we use the form of the GUP given by Eq.$%
\left( \ref{argup}\right) $ and investigate the thermodynamics of the
Schwarzschild black hole. We use units $\hbar =c=k_{ {B}}=G=1$.

\section{Massive Scalar field on the Randall-Sundrum brane Background}

We consider a dual-brane Randall-Sundrum scenario, embedded in a
5-dimensional {\it AdS}$_{5}$ spacetime. The 3-branes with
positive and negative tension are respectively located at the $S^{1}/Z_{2}$
orbifold fixed points $y=0$ and $y=y_{c}=\pi r_{c}$ \cite{sundrum1,sundrum2}%
. Assuming Poincar$\grave{e}$ invariance on the branes, the solutions
to Einstein's equations are given by,
\begin{equation}
ds^{2}=e^{-2ky}g_{\mu \nu }dx^{\mu }dx^{\nu }+dy^{2},
\end{equation}%
where the parameter $k$, assumed to be of the order of the Planck scale, governs the degree of curvature of the {\it AdS}$_5$ spacetime. Assuming a Ricci flat metric, one
solution is \cite{CHR}
\begin{equation}
ds^{2}=e^{-2ky}\left( -f(r)dt^{2}+f^{-1}(r)dr^{2}+r^{2}d\theta
^{2}+r^{2}sin^{2}\theta d\phi ^{2}\right) +dy^{2},  \label{metric}
\end{equation}%
where $f(r)=1-\frac{2M}{r}$. This solution describes a 4-dimensional
Schwarzschild black hole located on the hypersurface. It describes also a
5-dimensional  {\it AdS} \ black string intersecting the brane world.

Let us then consider a matter field propagation in this brane background. We
consider massive scalar field which are solutions of the Klein-Gordon
equation
\begin{equation}
(\nabla _{(5)}^{2}-m^{2})\Psi =0.  \label{kg}
\end{equation}%
Using the solution $\left( \ref{metric}\right) $ we have
\begin{eqnarray}
&&e^{2ky}\left[ -\frac{1}{f}\partial _{t}^{2}\Psi +\frac{1}{r^{2}}\partial
_{r}\left( r^{2}f\partial _{r}\Psi \right) +\frac{1}{r^{2}\mathrm{sin}\theta
}\partial _{\theta }(\mathrm{sin}\theta \partial _{\theta }\Psi )+\frac{1}{%
r^{2}\mathrm{sin}^{2}\theta }\partial _{\phi }^{2}\Psi \right]   \nonumber
\label{sol} \\
&&+e^{4ky}\partial _{y}(e^{-4ky}\partial _{y}\Psi )-m^{2}\Psi =0,
\end{eqnarray}%
Substituting $\Psi =e^{-i\omega t}\Phi (r,\theta ,\phi )\xi (y)$, we obtain
\begin{equation}\left( e^{ky_{c}}G(k,m )-G(0,m )\right)
\partial _{r}^{2}\Phi +\left( \frac{1}{f}\partial _{r}f+\frac{2}{r}\right)
\partial _{r}\Phi +\frac{1}{f}\left( {\frac{1}{r^{2}}}\left[ \partial
_{\theta }^{2}+\mathrm{cot}\theta \partial _{\theta }+{\frac{1}{\mathrm{sin}%
^{2}\theta }}\partial _{\phi }^{2}\right] +\frac{\omega ^{2}}{f}-\mu
^{2}\right) \Phi =0,  \label{rad}
\end{equation}%
where the constant $\mu ^{2}$ is defined by
\begin{equation}
e^{4ky}\partial _{y}(e^{-4ky}\partial _{y}\xi (y))-m^{2}\xi (y)+\mu
^{2}e^{2ky}\xi (y)=0.  \label{extra}
\end{equation}%
We simplify these equations by using the Wentzel-Kramers-Brillouin (WKB)
approximation for which we set $\Phi \sim e^{iS\left( r,\theta ,\phi \right)
}$. Indeed to leading order we have
\begin{eqnarray}\left( e^{ky_{c}}G(k,m )-G(0,m )\right)
-{\partial _{r}^{2}}\Phi -\left( \frac{1}{f}{\partial _{r}f}+\frac{2}{r}%
\right) \partial \Phi  &=&p_{r}^{2}\Phi , \\
-e^{4ky}\partial _{y}\left( e^{-4ky}\partial _{y}\xi (y)\right)
&=&p_{y}^{2}\xi (y),
\end{eqnarray}%
with $p_{\alpha }=\frac{\partial S}{\partial \alpha },\alpha =r,\theta ,\phi
,$ and $p_{r}^{2}$, $p_{y}^{2}$ given respectively by
\begin{eqnarray}
p_{r}^{2} &=&\frac{1}{f}\left( \frac{\omega ^{2}}{f}-\mu ^{2}-\frac{%
p_{\theta }^{2}}{r^{2}}-\frac{p_{\phi }^{2}}{r^{2} {sin}^{2}\theta }%
\right) , \\
p_{y}^{2} &=&\mu ^{2}e^{2ky}-m^{2}.
\end{eqnarray}%
A central ingredient for our calculation is the degeneracy of the brane and
extra dimension modes. To this aim, we first note that the volume in the momentum
space is affected by the squeezed momentum measure arising from the GUP and
given by Eq.$\left( \ref{ferm}\right) .$ Indeed, the number of quantum radial modes with energy less than $\omega $, for a given $\mu ,$ is given by
\begin{eqnarray}
n_{r}(\omega ) &=&\frac{1}{(2\pi )^{3}}\int drd\theta d\phi dp_{r}dp_{\theta
}dp_{\phi }e^{-\alpha p^{2}}   \nonumber \\
&=&\frac{2}{3\pi }\int dr\frac{r^{2}}{\sqrt{f}}\left( \frac{\omega ^{2}}{f}%
-\mu ^{2}\right) ^{3/2}e^{-\alpha \left( \frac{\omega ^{2}}{f}-\mu
^{2}\right) },  \label{nr}
\end{eqnarray}%
with the condition $\omega \geq \mu \sqrt{f}.$ We note that the
additional suppressing exponential, due to the GUP, renders
$n_{r}(\omega ) $ finite at the horizon without the introduction of
any artificial cut-off, as it is the case in the brick wall method

On the other hand, the
number of quantum states in the extra dimension for  given $\mu $ is%
\begin{eqnarray}
n_{y}\left( \mu \right)  &=&\frac{1}{\pi }\int dydp_{y}e^{-\alpha p_{y}^{2}}
 \nonumber \\
&=&\frac{1}{2\sqrt{\pi \alpha }}\int_{0}^{y_{c}} {  erf}\left( \sqrt{%
\alpha }\sqrt{\mu ^{2}e^{2ky}-m^{2}}\right) dy.  \label{em}
\end{eqnarray}%

\section{Entropy to all orders in the Planck length}

In this section, we shall evaluate the free energy and entropy of  free massive bulk
scalar fields at the Hawking temperature. We shall consider first the case where the GUP affect the bulk modes and finally the case where the GUP affect only the brane modes.
\paragraph{a) GUP on the bulk:}

In the continuum limit, the free energy of a scalar field at the inverse temperature $\beta$, is approximated by
\begin{equation}
F_{\beta }=\frac{1}{\beta }\int dN(\omega ) {ln}\left( 1-e^{-\beta
\omega }\right) .
\end{equation}%
where the total number of quantum states with energy less than $\omega $ is
given by
\begin{equation}
N(\omega )=\int ~dn_{r}~dn_{y}.  \label{N}
\end{equation}
A integration by parts gives
\begin{equation}
F_{\beta }=-\int_{\mu \sqrt{f(r)}}^{\infty }d\omega \frac{N(\omega )}{%
e^{\beta \omega }-1},
\end{equation}%
Using the expression of ${n_{r}}$ given by $(\ref{nr})$ we have
\begin{equation}
F_{\beta }=-\frac{2}{3\pi }\int_{r_h}^{r_{h}+\epsilon }dr\frac{r^{2}}{\sqrt{f}}%
\int_{m}^{\frac{\omega }{\sqrt{f(r)}}}d\mu \frac{dn_{y}\left( \mu \right) }{%
d\mu }g(\mu ),  \label{fe}
\end{equation}%
with
\begin{equation}
g(\mu )=\int_{\mu \sqrt{f(r)}}^{\infty }d\omega \left( \frac{\omega ^{2}}{%
f(r)}-\mu ^{2}\right) ^{3/2}\frac{e^{-\alpha \left( \frac{\omega ^{2}}{f(r)}%
-\mu ^{2}\right) }}{e^{\beta \omega }-1}.
\end{equation}%
Before proceeding further, we note that we are only interested in contributions to the entropy in the near vicinity of the horizon. Then, near horizon geometry considerations allows us to use the following
substitutions: $f\rightarrow 0$, $\frac{\omega ^{2}}{f}-\mu ^{2}\rightarrow
\frac{\omega ^{2}}{f}$, and then $g(\mu )$ is simply given by
\begin{equation}
g(\mu )=\frac{1}{f^{3/2}}\int_{0}^{\infty }d\omega \omega ^{3}\frac{e^{-%
\frac{\alpha \omega ^{2}}{f}}}{e^{\beta \omega }-1}.
\end{equation}%
Substituting in Eq.$\left( \ref{fe}\right) $ we obtain%
\begin{equation}
F_{\beta }=-\frac{2}{3\pi }\int_{r_{h}}^{r_{h}+\epsilon }dr\frac{r^{2}}{f^{2}%
}\int_{0}^{\infty }d\omega \omega ^{3}\frac{e^{-\frac{\alpha \omega ^{2}}{f}}%
}{e^{\beta \omega }-1}\int_{m}^{\frac{\omega }{\sqrt{f(r)}}}d\mu \frac{dn_{y}%
}{d\mu }.
\end{equation}%
At this stage the extra mode is completely decoupled from the radial modes
and it remains to integrate over $\mu $. Integrating over $y$ in Eq.$\left( \ref%
{em}\right) $ we obtain
\begin{equation}
n_{y}\left( \omega \right) =\frac{1}{2k\sqrt{\pi \alpha }}\int_{m}^{\frac{%
\omega }{\sqrt{f}}}\frac{d\mu }{\mu }\left(  {  erf}\left( \sqrt{\alpha }%
\sqrt{\mu ^{2}e^{2k\pi r_{c}}-m^{2}}\right) - {  erf}\left( \sqrt{\alpha }%
\sqrt{\mu ^{2}-m^{2}}\right) \right) .
\end{equation}%
The integration over $\mu $ can not be done exactly. To remedy  to this
situation we invoke the little mass approximation, for which we have the
following substitutions
\begin{equation}
\mu ^{2}e^{2k\pi r_{c}}-m^{2}\rightarrow \mu ^{2}e^{2k\pi r_{c}},\quad \mu
^{2}-m^{2}\rightarrow \mu ^{2}, { unless }\quad\mu =m.  \label{app}
\end{equation}%
Then the free energy is rewritten as
\begin{equation}
F_{\beta }=-\frac{2}{3\pi }\int_{r_{h}}^{r_{h}+\epsilon }dr\frac{r^{2}}{f^{2}%
}I(r),  \label{f40}
\end{equation}%
where $I\left( r\right) $ is given by
\begin{equation}
I(r)=\frac{1}{2k\pi ^{3/2}\sqrt{\alpha }}\int_{0}^{\infty }d\omega \omega
^{3}\frac{e^{-\frac{\alpha }{f}\omega ^{2}}}{e^{\beta \omega }-1}\int_{m}^{%
\frac{\omega }{\sqrt{f}}}\frac{d\mu }{\mu }\left(  {  erf}\left( \sqrt{%
\alpha }\mu e^{k y_{c}}\right) - {  erf}\left( \sqrt{\alpha }\mu
\right) \right) .
\end{equation}%
The entropy is calculated using the first law of thermodynamics $S=\beta^2\frac{\partial F}{\partial\beta}$ as
\begin{equation}
S=\frac{4\beta^2}{3k\pi^{3/2}\alpha^{1/2}}\int_{r_{h}}^{r_{h}+\epsilon }dr\frac{r^{2}}{f^{2}%
}\int_{0}^{\infty}d\omega\omega^4\frac{e^{-\frac{\alpha}{f}\omega^2}}
{\hbox{sinh}^2(\beta\omega/2)}\int_{m}^{%
\frac{\omega }{\sqrt{f}}}\frac{d\mu }{\mu }\left(  {  erf}\left( \sqrt{%
\alpha }\mu e^{k y_{c}}\right) - {  erf}\left( \sqrt{\alpha }\mu
\right) \right) .
\end{equation}
In terms of the variable $x=\omega \sqrt{\alpha}$ we write the entropy as
\begin{equation}
S=\frac{4\beta^2}{3k\pi^{3/2}\alpha^{3}}\int_{0}^{\infty}dx\frac{x^4}{\hbox{sinh}^2(x\beta/2\sqrt{\alpha})}I(x,\epsilon),
\label{entropy}
\end{equation}
where  $I(x,\epsilon)$ is given by
\begin{equation}
I(x,\epsilon)=\int_{r_{h}}^{r_{h}+\epsilon }dr\frac{r^{2}}{f^{2}}e^{-\frac{x^2}{f}}\int_{m}^{%
\frac{x}{\sqrt{\alpha f}}}\frac{d\mu }{\mu }\left(  {  erf}\left( \sqrt{%
\alpha }\mu e^{k y_{c}}\right) - {  erf}\left( \sqrt{\alpha }\mu
\right) \right) .
\end{equation}
Now the integration over $\mu $ can be done exactly and we obtain
\begin{equation}
I(x,\epsilon)=2\sqrt{\frac{\alpha}{\pi}}\left( \frac{x}{\sqrt{\alpha}}I_0(x,\epsilon)-mI_m(x,\epsilon)\right),
\end{equation}
which is the sum of independent and dependent mass contributions given respectively by
\begin{equation}
I_0(x,\epsilon)=\int_{r_{h}}^{r_{h}+\epsilon }dr\frac{r^{2}}{f^{5/2}}e^{-\frac{x^2}{f}}\left[  \left( e^{ky_{c}}G(k,\frac{x}{\sqrt{\alpha f}} )-G(0,\frac{x}{\sqrt{\alpha f}} )\right) %
\right] ,
\end{equation}%
\begin{equation}
I_m(x,\epsilon)=\int_{r_{h}}^{r_{h}+\epsilon }dr\frac{r^{2}}{f^{2}}e^{-\frac{x^2}{f}}\left[  \left( e^{ky_{c}}G(k,m )-G(0,m )\right) %
\right],
\end{equation}
and where $G(k,\mu )$ is the hypergeometric function
\begin{equation}
G(k,\mu )=_{2}F_{2}\left( \frac{1}{2},\frac{1}{2};\frac{3}{2},\frac{3}{2}%
;-\alpha \mu ^{2}e^{2ky_{c}}\right) .
\end{equation}%
Before proceeding any further, let us carefully analyze the integration over $r$. Because of the near horizon considerations
we have, to order ${\cal{O}}\left((r-r_h)^2\right)$, the following approximation
\begin{equation}
f(r)\simeq (r-r_h)\frac{df}{dr}\vert_{r_h}=2\kappa(r-r_h),
\end{equation}
where $\kappa=2\pi/\beta$ is the surface gravity at the horizon.
Now we proceed to the calculation of $I_0$ and $I_m$. We first write $I_0$ as 
\begin{equation}
I_0(x,\epsilon)=\sum_{n=0}^{\infty}
(-1)^n\frac{a_n^2\gamma_n}{n!}\int_{r_{h}}^{r_{h}+\epsilon }dr\frac{r^{2}}{(2\kappa(r-r_h))^{5/2}}{\left(\frac{x^2}{2\kappa(r-r_h)}\right)^n}e^{-\frac{x^2}{2\kappa(r-r_h)}}
\end{equation}
where $a_n=\frac{(1/2)_n}{(3/2)_n}$ and $(z)_n=\frac{\Gamma(n+z)}{\Gamma(z)}$ is the Pochhammer symbol, and $\gamma_n=e^{(2n+1)ky_c}-1$.
With the change of variable $t=\frac{x^2}{2\kappa(r-r_h)}$, $I_0$ becomes
\begin{equation}
I_0(x,\epsilon)=\frac{1}{2\kappa}\sum_{n=0}^{\infty}
(-1)^n\frac{a_n^2\gamma_n}{n!}\int_{x^2/2\kappa\epsilon}^{\infty}\left(\frac{r_h^2}{x^3}+\frac{x}{4\kappa^2t^2}+\frac{r_h}{\kappa x t}\right)t^{n+1/2}e^{-t}dt.
\end{equation}
Using the definition of the incomplete Gamma function 
\begin{equation}
\Gamma(a,z)=\int_{z}^{\infty}t^{a-1}e^{-t}dt,
\end{equation}
we obtain
\begin{equation}
I_0(x,\epsilon)=\frac{1}{2\kappa}\sum_{n=0}^{\infty}
(-1)^n\frac{a_n^2\gamma_n}{n!}\left[\frac{r_h^2}{x^3}\Gamma(n+\frac{3}{2},\frac{x^2}{2\kappa\epsilon})
+\frac{x}{4\kappa^2}\Gamma(n-\frac{1}{2},\frac{x^2}{2\kappa\epsilon})
+\frac{2r_h}{x }\Gamma(n+\frac{1}{2},\frac{x^2}{2\kappa\epsilon})\right]. \label{I0}
\end{equation}
Repeating the same procedure for $I_m(x,\epsilon)$, we obtain
\begin{equation}
I_m(x,\epsilon)=\left( e^{ky_{c}}G(k,m )-G(0,m )\right)\left[\frac{r_h^2}{2\kappa x^2}e^{-x^2/2\kappa\epsilon}
+\frac{x^2}{8\kappa^3}\Gamma(-1,\frac{x^2}{2\kappa\epsilon})+\frac{r_h}{2\kappa^2}\Gamma(0,\frac{x^2}{2\kappa\epsilon})\right].
\label{I1}
\end{equation}
At this stage the brick wall cutoff $\epsilon$ can be  related in our framework to the physical scale given by the minimal length as
\begin{equation}
 \left(\delta X\right)_0=\int_{r_h}^{r_h+\epsilon}\frac{dr}{\sqrt{f(r)}}.
\end{equation}
This relation gives
\begin{equation}
\epsilon=\frac{e\kappa\alpha}{4}.
\end{equation}
Then using this expression in $(\ref{I0})$ and $(\ref{I1})$ and substituting in $(\ref{entropy})$ we obtain the final expression of the near horizon entropy
\begin{eqnarray}
S&=&\frac{8e}{3k\pi^3\alpha^{1/2}}\bigg(\gamma_1\left(a_0\frac{A}{A_0}+\frac{b_0}{4\pi^4e^2}\frac{A_0}{A}+\frac{c_0}{\pi^2e}\right)\nonumber\\
&-&
\frac{\gamma_3}{9}\left(a_1\frac{A}{A_0}+\frac{b_1}{4\pi^4e^2}\frac{A_0}{A}+\frac{c_1}{\pi^2e}\right)\bigg)\nonumber\\
&-&\frac{8em}{3k\pi^3}\left( e^{ky_{c}}G(k,m )-G(0,m )\right)
\left(a_2\frac{A}{A_0}
+\frac{b_2}{4\pi^4 e^2}\frac{A_0}{A}+\frac{c_2}
{\pi^2 e}\right),\label{entropy1}
\end{eqnarray}
where $A=4\pi r_{h}^{2}$ , $A_{0}=4\pi \left( \delta X\right) _{0}^{2}$ is the minimal black hole area due to the GUP, and 
the numerical values $a_i,b_i,c_i (i=1,2,3)$ are given by
\begin{eqnarray}
\int_0^{\infty}dy\frac{y^2}{\hbox{sinh}^2(y)}\Gamma(a,\frac{2y^2}{\pi^2 e})=
\begin{cases}
a_0\simeq 1.2195 &\text{for } a=3/2\\
a_1\simeq 2.0382 &\text{for } a=5/2
\end{cases}
\end{eqnarray}
\begin{eqnarray}
\int_0^{\infty}dy\frac{y^6}{\hbox{sinh}^2(y)}\Gamma(a,\frac{2y^2}{\pi^2 e})=
\begin{cases}
b_0\simeq 12.1968 &\text{for } a=-1/2\\
b_1\simeq  9.3742 &\text{for } a=1/2\\
b_2\simeq  18.4608 &\text{for } a=-1
\end{cases}
\end{eqnarray}
\begin{eqnarray}
\int_0^{\infty}dy\frac{y^4}{\hbox{sinh}^2(y)}\Gamma(a,\frac{2y^2}{\pi^2 e})=
\begin{cases}
c_0\simeq 2.2912 &\text{for } a=1/2\\
c_1\simeq 2.9991 &\text{for } a=3/2 \\
c_2\simeq 3.0706 &\text{for } a=0
\end{cases}
\end{eqnarray}
and 
\begin{equation}
a_2=\int_{0}^{\infty}dy\frac{y^2}{\hbox{sinh}^2(y)}e^{-\frac{2y^2}{\pi^2 e}}\simeq 1.4508
\end{equation}
We note that the mass independent contribution to the entropy is just built from the two first terms of $I_0$, since the factors  of the type $(a_n)^2/n!$ become small for $n\geq2$.
Some comments are appropriate
about the the expression of the entropy given by ($\ref{entropy1}$).
It is interesting to note that the entropy shows two regimes. In a first regime of weak gravitational fields corresponding to $\alpha$ small, we have the usual Bekenstein-Hawking area law $S\thicksim{A/A_0}$, while in the second regime of strong gravitational field corresponding to large values of $\alpha$, the entropy bahaves like $S\thicksim{A_0/A}$. However, the constraint $A\geq A_0$ imposed by the GUP, renders the correction term small and the Bekentein-Hawking term is the dominant contribution to the entropy. We note, that corrections to the horizon area law of the entropy of the Schwarzschild black hole in the ADD scenario with GUP have been obtained recently \cite{nouicer4}. These deviations from the horizon area law have not been obtained in some recent works without GUP \cite{medved} and with a GUP to leading order in the Planck length \cite{kim}.
Finally we note, that our result has been obtained with the aid of the little mass approximation , and due to
the existence of a minimum black hole area, it is non-perturbative in the minimal length. On the other hand the
massive correction contribution is more complicated than the one obtained in \cite{medved,kim}, where it is linear
in $m$. 

\paragraph{b) GUP on the brane:}
We consider now the more interesting case where the modes in the extra dimension are
not affected by the GUP. In such a situation the number of quantum extra modes
is simply given  by
\begin{equation}
n_{y}=\frac{1}{\pi }\int_{0}^{y_{c}}\sqrt{\mu ^{2}e^{2ky}-m^{2}}dy
\end{equation}
and the total number with energy less than $\omega $ is
\begin{equation}
n_{y}\left( \omega \right) =\frac{1}{k\pi }\int_{m}^{\frac{\omega }{\sqrt{f}}%
}\frac{d\mu }{\mu }\left( \sqrt{\mu ^{2}e^{2ky_{c}}-m^{2}}-\sqrt{\mu
^{2}-m^{2}}\right) .
\end{equation}%
The calculation of the free energy proceeds as in the previous section and
is given by
\begin{equation}
F_{\beta }=-\frac{2}{3k\pi^2 }\int_{r_h}^{r_{h}+\epsilon }dr\frac{r^{2}}{f^{2}}\int_{0}^{\infty }d\omega \omega ^{3}\frac{e^{-\frac{%
\alpha }{f}\omega ^{2}}}{e^{\beta \omega }-1}\int_{m}^{\frac{\omega }{\sqrt{f%
}}}\frac{d\mu }{\mu }\left( \sqrt{\mu ^{2}e^{2k\pi r_{c}}-m^{2}}-\sqrt{\mu
^{2}-m^{2}}\right).
\label{fe2}
\end{equation}%
The entropy is calculated from the relation $S=\beta^2\partial F/\partial \beta$. In terms of the variable $x=\omega\sqrt{\alpha}$ and $z=\mu/m$ we have
\begin{equation}
S=\frac{2\beta^2m}{3k\pi^2\alpha^{5/2}}\int_{r_{h}+\epsilon }dr\frac{r^{2}}{f^{2}}\int_{0}^{\infty}dx\frac{x^4 e^{-x^2/f}}{\hbox{sinh}^2
(\beta x\/2\sqrt{\alpha})}J\left( x\right),
\end{equation}%
with $J\left( x\right) $ is given by
\begin{equation}
J\left( x\right) =\int_{1}^{\frac{x}{m\sqrt{\alpha f}}}\frac{dz}{z}\left(
\sqrt{z^{2}e^{2k\pi r_{c}}-1}-\sqrt{z^{2}-1}\right) .
\end{equation}%
The integration over $z$ is straightforward, and as a result we obtain%
\begin{eqnarray}
J\left( x\right) &=&\sqrt{\left( \frac{xe^{ky}}{m\sqrt{\alpha f}}\right)
^{2}-1}+\arctan \left( \frac{1}{\sqrt{\left( \frac{xe^{ky}}{m\sqrt{\alpha f}}%
\right) ^{2}-1}}\right)   \nonumber \\
&&-\sqrt{\left( \frac{x}{m\sqrt{\alpha f}}\right) ^{2}-1}-\arctan \left(
\frac{1}{\sqrt{\left( \frac{x}{m\sqrt{\alpha f}}\right) ^{2}-1}}\right)
 \nonumber \\
&&-\sqrt{e^{2ky}-1}-\arctan \left( \frac{1}{\sqrt{e^{2ky}-1}}\right) +\frac{%
\pi }{2}.
\end{eqnarray}%
In the just vicinity of the horizon, corresponding to $f\rightarrow 0$, we
have the approximation
\begin{equation}
J\left( x\right) \approx \frac{x}{m\sqrt{\alpha }}\gamma _{1}-\left( \sqrt{%
\gamma _{2}}+\arctan \left( \frac{1}{\sqrt{\gamma _{2}}}\right) \right) .
\label{xf}
\end{equation}%
where $\gamma _{a}=e^{aky_{c}}-1$. 

Then the entropy can be written as
\begin{equation}
S=S_0+S_m,
\end{equation}%
where
\begin{equation}
S_0=\frac{2\beta^2\gamma_1}{3k\pi^2\alpha^3}\int_{0}^{\infty}dx\frac{x^5 }{\hbox{sinh}^2
(\beta x/2\sqrt{\alpha})}\int_{r_h}^{r_{h}+\epsilon }dr\frac{r^{2}e^{-x^2/f}}{f^{5/2}},
\end{equation}
and
\begin{equation}
S_m=\frac{2\beta^2m}{3k\pi^2\alpha^{5/2}}\left(\sqrt{\gamma_2}+\text{tan}^{-1}\left(\frac{1}{\sqrt{\gamma_2}}\right)\right)\int_{0}^{\infty}dx\frac{x^4 }{\text{sinh}^2
(\beta x/2\sqrt{\alpha})}\int_{r_h}^{r_{h}+\epsilon }dr\frac{r^{2}e^{-x^2/f}}{f^{2}}.
\end{equation}
Following the same steps of calculation as in the first case, the integrals about $r$ are computed and we obtain the final expression of the entropy
\begin{eqnarray}
S&=&\frac{2 e\gamma_1}{3k\pi^3\alpha^{1/2}}\left(a_2\frac{A}{A_0}+\frac{b_0}{4\pi^4 e^2}\frac{A_0}{A}+\frac{c_0}{\pi^2 e}\right)\nonumber\\
&-&\frac{2 em}{3k\pi^3}\left( \sqrt{%
\gamma_{2}}+\arctan \left( \frac{1}{\sqrt{\gamma _{2}}}\right) \right)\left(a_2\frac{A}{A_0}+\frac{b_2}{4\pi^4 e^2}\frac{A_0}{A}+\frac{c_2}{\pi^2 e}\right),\label{entropy2}
\end{eqnarray}
where the numerical constants are given by Eqs.(56-59).

We note that the entropy given by ($\ref{entropy2}$) exhibits the same two regimes noted in the case where the GUP is applied on the full volume of the spacetime.  We observe that the mass contribution to the entropy becomes linear as obtained in \cite{medved,kim}. This a consequence of the suppression of the damping of the states density in the extra dimension direction.

Before ending this section, let us comment about the  entropy to all orders in the Planck length for the (3+1)-dimensional Schwarzschild black hole obtained in \cite{park} and given by
\begin{equation}
S=\frac{e^3\zeta(3)}{8\pi\alpha}A.\label{par}
\end{equation}
However, following the procedure developped in this section, the evaluation of the integral over $r$ in the range near horizon gives
\begin{equation}
 S=\frac{ea_2}{6\pi^2}\frac{A}{A_0}+\frac{b_2}{24\pi^6 e}\frac{A_0}{A}+\frac{2c_2}{3\pi^3},
\end{equation}
where $a_2,b_2,c_2$ are given above. In comparison with Eq.$(\ref{par})$, our result shows again the small deviation from the Bekenstein-Hawking area law, proportional to the inverse of the horizon area.
Finally we point that, even with a GUP to leading order in the Planck length, a careful evaluation of the entropy integrals about $r$ in the range near horizons of the Randall-Sundrum black brane shows the same small correction terms to the Bekenstein-Hawking area law obtained in \cite{kim}.

\section{Conclusion}

In summary, we have calculated to all orders in the Planck length,
the near horizon contributions to the entropy of bulk massive scalar
fields propagating in the background of a black hole in the
Randall-Sundrum brane world, by using the generalized
uncertainty principle. The entropy is obtained by summing up the
thermal contributions of both the brane and the extra dimension fields. As a result the usual Bekenstein area law is 
not preserved in our framework and is corrected by the a term proportional to the inverse of the horizon area. Our analysis shows that the usual Bekenstein area term remains the dominant contribution since by virtue of the GUP, the correction term relevant in the case of strong gravitational fields, is a small quantity. In the case when the GUP is
considered on the full volume of the bulk, we have shown
that the mass dependence of the entropy is more complicated in
comparison to the linear mass contribution obtained in refs.
\cite{medved} and \cite{kim}. The later behavior is recovered when
the effect of the GUP in the extra dimension direction is ignored. As a consequnce the masive contribution to the entropy depends crucialy on the presence or not of a cutoff in the extra dimension direction. Finally, we
note that the results obtained are non perturbative in the minimal
length.

{\textbf{Acknowledgments}}: The author thanks the Algerian Ministry
of Scientific Research and High Education for financial support and
the Professor Walter Greiner from FIAS-Frankfurt for warm hospitality.

\end{document}